\documentclass[%
 aip,
 amssymb,amsmath,amsfonts,mathrsfs,bm,gensymb,
 reprint,%
]{revtex4-1}

\usepackage{graphicx}
\usepackage{dcolumn}
\usepackage{bm}

\usepackage[utf8]{inputenc}
\usepackage[T1]{fontenc}
\usepackage{mathptmx}
\usepackage{etoolbox}
\usepackage{mathtools}
\usepackage{hyperref}

\usepackage{color}

\hypersetup{
    colorlinks=true,
    linkcolor=black,
    citecolor=black,
    filecolor=black,
    urlcolor=black,
}

\makeatletter
\def\@email#1#2{%
 \endgroup
 \patchcmd{\titleblock@produce}
  {\frontmatter@RRAPformat}
  {\frontmatter@RRAPformat{\produce@RRAP{*#1\href{mailto:#2}{#2}}}\frontmatter@RRAPformat}
  {}{}
}%
\makeatother
\begin{document}

\preprint{AIP/123-QED}

\title[The detection matrix as a model-agnostic tool in mechanics]{The detection matrix as a model-agnostic tool to estimate the number of degrees of freedom in { mechanical systems and engineering structures}}
\author{Paolo Celli}
 \email{paolo.celli@stonybrook.edu, mporfiri@nyu.edu}
 \affiliation{Department of Civil Engineering, Stony Brook University, Stony Brook, NY 11794, USA}
 
\author{Maurizio Porfiri}%
\affiliation{Center for Urban Science and Progress, Tandon School of Engineering, New York University, NY 11201, USA}%
\affiliation{Department of Biomedical Engineering, Tandon School of Engineering, New York University, NY 11201, USA}
\affiliation{Department of Mechanical and Aerospace Engineering, Tandon School of Engineering, New York University, NY 11201, USA}

\date{\today}

\begin{abstract}
\vspace{5px}
\normalsize{\textbf{This article may be downloaded for personal use only. Any other use requires prior permission of the author and AIP Publishing. This article appeared in}: \emph{Chaos} {\bf 32}, 033106 (2022) \textbf{and may be found at}: \url{https://doi.org/10.1063/5.0083767}}
\vspace{15px}

Estimating the number of degrees of freedom of a  {mechanical system or an engineering structure} from the time-series of a small set of sensors is a basic problem in diagnostics, which, however, is often overlooked when monitoring their health and integrity. In this work, we demonstrate the applicability of the network-theoretic concept of detection matrix as a tool to solve this problem. From this estimation, we illustrate the possibility to identify damage. The detection matrix, recently introduced by Haehne \textit{et al.} in the context of network theory~\cite{Haehne2019}, is assembled from the transient response of a few nodes as a result of non-zero initial conditions: its rank offers an estimate of the number of nodes in the network itself. The use of the detection matrix is completely model-agnostic, whereby it does not require any knowledge of the system dynamics. Here, we  {show} that, with a few modifications, this same principle applies to discrete systems, such as spring-mass lattices and trusses. Moreover, we discuss how damage in one or more members causes the appearance of distinct jumps in the singular values of this matrix, thereby opening the door to structural health monitoring applications, without the need for a complete model reconstruction.
\end{abstract}

\maketitle

\begin{quotation}
The design of  {civil} structures has been a pillar of human civilization since its very origins. Egyptian pyramids and Roman aqueducts are some of the most fascinating, large structures that humankind has created to date. How many stones make up a pyramid? How many bricks are still intact in the ruins of an aqueduct? Are any of these structures at risk of collapsing?  {Analogous question arise in mechanical and aerospace engineering and can potentially} be answered by looking beyond the boundaries of  {mechanics}, drawing inspiration from the field of network theory. Through the network-theoretic concept of a detection matrix, it is possible to count the total number of degrees of freedom in  {these structures} from the measurement of the dynamic response of some of  {their} units that will contain a footprint of  {all the other units}. Ideally, we could count the stones in a pyramid or the bricks in an aqueduct by measuring the vibration of a few of them across several experiments where we would mechanically excite different parts of the structure. From the detection matrix, we can also gather insight into the health of the structure, pinpointing potential damage that could ultimately trigger failure.  {This study makes a preliminary step in this direction, opening the door for new synergies between  {mechanics} and network theory, two fields that have remained disjoint for too long.}
\end{quotation}

\section{Introduction}

Monitoring the health of  {mechanical systems and engineering structures} is of utmost importance in civil, aerospace, and mechanical engineering. In general, the onset of damage is associated with modifications to global properties with respect to a baseline undamaged state~\cite{Doebling1998}. The most commonly used properties for damage identification are those associated with dynamic response: stiffness, mass, and damping~\cite{Salawu1997}. This is due to the feasibility of \textit{in-situ} dynamic tests on mechanical systems and engineering structures through common and low-cost sensors and readily available excitation signals. Accelerometers, strain gauges, and laser displacement sensors are often employed to sense mechanical vibration at desired locations of a structure. Excitation signals are either generated by external means, such as impact hammers or shakers, or are part of the normal operation of the  {system}~\cite{Avci2021}. 

Over the years, many vibration-based structural health monitoring (SHM) methods have been introduced for applications  {in mechanical systems and engineering structures}~\cite{Doebling1998}. Classical methods rely on the detection of the signatures of damage from modal parameters such as natural frequencies, mode shapes, and modal damping~\cite{Doebling1998, Avci2021}. A common drawback of these methods is their sensitivity to confounding factors that could mask the effect of damage on modal parameters, such as temperature and humidity~\cite{Deraemaeker2008, Ubertini2017}. Another limitation is that these methods often rely on detailed finite element models of the  {system} being investigated to correctly identify its properties and detect the onset of damage~\cite{Doebling1998}. Finally, despite few recent advances~\cite{Yang2013, Hosseinzadeh2014, Bagheri2017}, most of these methods rely on a large number of sensors and vast amounts of data, both difficult to attain in real applications. 

To overcome the limitations of classical vibration-based methods, recent efforts have concentrated on data-driven and model-agnostic SHM techniques~\cite{Avci2021}. A first example is represented by techniques based on the statistical analysis of time-series~\cite{KOPSAFTOPOULOS20101977}. Based on the extraction of features from the data and on the comparison between these features at different points throughout the life of a  {system}~\cite{Nair2006, Gul2009, Shokravi2020}, these techniques do not require the knowledge of the system parameters~\cite{Fassois2007}. Among them, particularly relevant to our work are those based on subspace identification and on the creation of a block Hankel matrix of output data~\cite{Deraemaeker2008}. Another family of data-driven techniques that is gaining considerable attention is based on artificial intelligence and machine learning~\cite{SALEHI2018170, Avci2021}. Such techniques are based on algorithms that can be trained to recognize novelty in the response of a system, that is, to identify if its characteristics have changed due to damage~\cite{Figueiredo2011, Druce2015, neves2017structural}. 
A less studied area, which still holds great promise is the use of information-theoretic tools in which the presence of damage is associated with a modification of the flow of information within the  {system}, from a source to a receiver 
\cite{sudu2016information}.
Despite their many advantages, existing methods rely on the collection of large amounts of data, thereby posing challenges at the levels of sensor placement and data transfer~\cite{SALEHI2018170}.
 
New pathways to address some of the challenges in SHM, or to propose alternatives to the existing methods, can be found by recognizing analogies between  {mechanics and network theory}. For example, a one-dimensional (1D) mass-spring lattice can be viewed as a path graph in which springs define the edges of the network and the masses behave as nodes. Likewise, trusses and frames can be assimilated to complex networks, with nodal dynamics unfolding in multiple dimensions corresponding to the degrees of freedom of the joints~\cite{Shai1999}. Recently, network-theoretic tools have been used to identify damage in 2D disordered beam lattices~\cite{Berthier2019},  demonstrating the power of analogies  {between mechanics and network theory}~\cite{Moretti2019}. In particular, Berthier \textit{et al.}~\cite{Berthier2019} resorted to the concept of geodesic edge betweenness centrality to identify which beams are most likely to fail in response to applied static loads; this idea was, in turn, derived from previous studies based on the analogy between granular materials and networks~\cite{Smart2008, Kollmer2019}. 

Analogies between  {engineering structures} and networks can also be drawn based on their dynamic behavior. Historically, similarities between structural dynamics and complex networks have been at the heart of optimization methods, thereby allowing for simultaneous optimization of topology, geometry, and sizing of trusses~\cite{giger2006evolutionary}, as well as allowing to perform exhaustive analyses of the entire topology space in search of constrained global optima \cite{kawamoto2004planar}. These analogies have been fueled by the availability of a variety of mathematical tools from network theory on which to base rigorous analysis. 

However, the field of complex networks has witnessed a profound transformation over the last decade, with research focus gradually shifting from analysis to diagnostics \cite{runge2019inferring,pilkiewicz2020decoding,boers2021complex,bullmore2009complex}. To date, the network community has put forward a wide range of methodologies to undertake the problem of inferring key properties of a network from partial knowledge of the dynamics of its nodes. By sampling the time-series of the nodal dynamics, one can now reconstruct the topology of networks and pinpoint cause-and-effect relationships within the system. For example, it is now possible to isolate leader-follower relationships in collective behavior of animal groups from raw time-series of individual motion~\cite{pilkiewicz2020decoding}, determine causal networks of interactions in Earth science from observational data~\cite{runge2019inferring,boers2021complex}, and infer functional connectivity patterns in the brain from electroencephalography or other measurement techniques~\cite{bullmore2009complex}.

\begin{figure*}[!htb]
\centering
\includegraphics[scale=1.1]{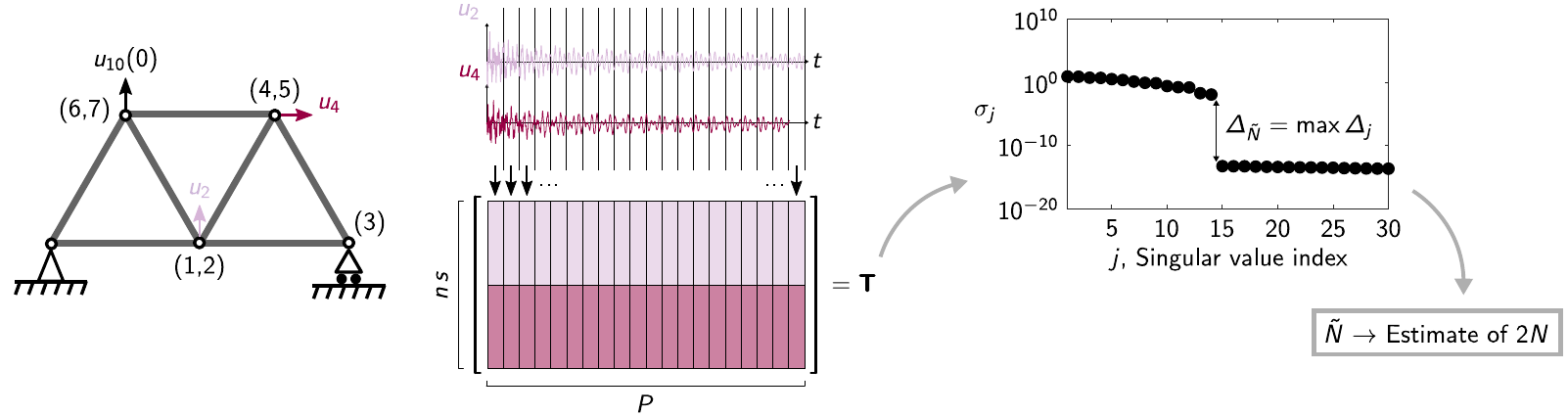}
\caption{Schematic of the method, that illustrates how the detection matrix $\mathbf{T}$ is populated from data recorded for $n$ time traces in response to initial conditions on $N_\mathrm{ic}$ degrees of freedom. The response signal, comprising $S$ time instants, is divided into $P$ segments, each composed of $s=\lfloor S/P \rfloor$ measurements. We call $N$ the number of degrees of freedom in the system. $\tilde{N}$, the index of the singular value $\sigma_j$ of $\mathbf{T}$ corresponding to the maximum spectral gap $\Delta_j$, represents an estimate of $2N$.}
\label{f:meth}
\end{figure*}

The field of mechanics, for the most part, has yet to seize the opportunity to capitalize on these efforts. In this work, we make a  {first, preliminary step} in this direction by addressing what is arguably the most fundamental step in any diagnostics problem, that is, counting the number of elements of a  {mechanical system or an engineering structure} by using time-series from the motion of only a small portion of the  {system} itself. More concretely, suppose we are given the 2D truss shown in Fig.~\ref{f:meth} with seven degrees of freedom, but we only have partial measurements of the motion of one or two of its degrees of freedom. Can we then recover information on the unmeasured degrees of freedom and estimate the correct total number of degrees of freedom for the entire  {system}?

We provide a positive answer to this inference problem by adapting the methodology proposed by Haehne \textit{et al.}~\cite{Haehne2019} {for the study of nonlinear networked dynamical systems}, which is based on the use of a detection matrix. To some extent, the premise of this methodology of using  
{sampled time traces is similar to subspace identification techniques based on the block Hankel matrix~\cite{van2012subspace}}.
For a network of first-order continuous-time dynamical systems, the detection matrix collates the time-series of some of the network nodes across multiple, independent experiments. As  {analytically demonstrated} by Porfiri~\cite{Porfiri2020} under quite general conditions, one can infer the total number of nodes in the network without any knowledge of the network itself from the rank of the detection matrix. Although the mathematical  {derivation} relies on the linearity of the networked dynamical system, complex time-varying dynamics can be considered in a continuous-time settings, unlike the typical settings of subspace identification methods that are based on time-invariant discrete-time models~\cite{van2012subspace}.

Here, we {  demonstrate the application of the detection matrix to tackle problems related to the dynamic response of  {mechanical systems and engineering structures}, governed by second order dynamics. To ensure the practical viability of the approach, we construct the detection matrix from a single observation.} This method essentially allows one to detect the number of members in a  {system} by imparting a random initial condition at a single location, for instance, through the use of an impact hammer, and by measuring the time evolution of a very small subset of its available degrees of freedom, as schematically illustrated in Fig.~\ref{f:meth}, without any knowledge of the  {system} itself. 

Interestingly, this method thrives in the presence of uncertainties, and we demonstrate that random perturbations of the  {system} properties improve its efficiency. We also illustrate that the method performs well even under low levels of noise  {and in the presence of moderate nonlinearities}. With some modifications, the proposed method could be leveraged to detect the onset of damage in discrete  {systems} including trusses, frames, and other beam lattices. Unlike techniques based on subspace identification from output-only data~\cite{Deraemaeker2008, Cancelli2020}, this method does not rely on a complete identification of the system at hand -- but only on counting the number of degrees of freedom and on monitoring changes to this quantity. In a way, our method is similar in spirit to others that are based on the monitoring of a damage-sensitive feature of measured data~\cite{Dohler2013, Gres2021, Viefhues2022}, with the difference that our damage-sensitive feature, the number of degrees of freedom, has a clear physical meaning and is not affected by changes in the amplitude of the initial condition or in the environmental conditions. Its model-agnostic nature, however, makes it challenging if not impossible to detect where damage has occurred. Thus, this technique could be powerful at the early stages of a diagnostics problem to determine if damage is present, before any attempt to localize it.

The rest of the paper is organized as follows. In Section~\ref{s:theory}, we present the theory of the detection matrix, illustrate its algorithmic implementation for  {mechanical} problems, and detail the relationship between the detection matrix and the block Hankel matrix that is more familiar to the system identification community. In Section~\ref{s:dem}, we demonstrate the application of the method for the study of 1D spring-mass-damper lattices. Therein, we systematically assess the performance of the algorithm as a function of the location of the excitation and observation nodes, variations in physical properties (including cyclic variations), number of observation nodes, duration and noise of the time-series, and  {nonlinearities in the system response}. In this Section, we also summarize preliminary results on health monitoring. In Section~\ref{s:truss}, we extend the analysis to the study of trusses, whereby we demonstrate the viability of the approach to tackle 2D systems of varying complexity. Concluding remarks and an outlook on future research avenues are summarized in Section~\ref{s:con}. 

\section{Theory}
\label{s:theory}

{
\subsection{Derivation of the detection matrix for continuous-time, second-order linear systems}}
\label{s:Ttheory}

We consider a  {system} with $N$ degrees of freedom (DOFs) without external forces, described by general linear, time-invariant dynamics of the form~\cite{caughey1960classical, Meirovitch2010}
    \begin{equation}\label{eq:eqmot}
        \mathbf{M}\ddot{\mathbf{q}}(t)+\mathbf{D}\dot{\mathbf{q}}(t)+\mathbf{K}\mathbf{q}(t)=\mathbf{0}_N,
    \end{equation}
where $t\in\mathbb{R}_{\geq 0}$ is time, $\mathbf{q}\in\mathbb{R}^N$ is the vector of generalized coordinates, $\mathbf{M}\in\mathbb{R}^{N\times N}$ is the mass matrix, $\mathbf{D}\in\mathbb{R}^{N\times N}$ is the damping matrix, $\mathbf{K}\in\mathbb{R}^{N\times N}$ is the stiffness matrix, and $\mathbf{0}_N$ is the zero vector in $\mathbb{R}^N$. The three matrices are assumed to be symmetric and positive definite; this premise is in line with the standard treatment of linear vibrations of passive systems, although not strictly needed for the development of the theory that only requires the mass matrix to be non-singular. Instances in which we will exploit the properties of the mass, stiffness, and damping matrices largely pertain to the interpretation of the hypotheses and implications of our claims.

For convenience, we express \eqref{eq:eqmot} as a first order system in $\mathbb{R}^{2N}$ so that
    \begin{equation}\label{eq:LTI}
    \dot{\mathbf{x}}=\mathbf{A}{\mathbf{x}},
    \end{equation}
where $\mathbf{x}=\left[\mathbf{q}^{\mathrm{T}},\dot{\mathbf{q}}^{\mathrm{T}}\right]^{\mathrm{T}}$ and $\mathbf{A}\in \mathbb{R}^{2N\times 2N}$ is defined by
    \begin{equation}
        \mathbf{A}=\begin{bmatrix}
    \mathbf{0}_{N\times N}  &  \mathbf{I}_{N\times N}      \\
    -\mathbf{M}^{-1}\mathbf{K}  &  -\mathbf{M}^{-1}\mathbf{D}       
\end{bmatrix},
    \end{equation}
with ``$\mathrm{T}$'' being the matrix transposition, $\mathbf{I}_{N\times N}$ the identity matrix in $\mathrm{R}^{N\times N}$, and $\mathbf{0}_{N\times N}$ the zero matrix in $\mathrm{R}^{N\times N}$.

We assume to have a access to only a small subset of the DOFs, which may comprise one or more of the generalized coordinates and/or their time derivative. Hence, we measure the vector $\mathbf{y}\in\mathbb{R}^n$, where the number $n$ is the total number of available scalar time traces. In principle, these time traces could even be linearly combined such that
    \begin{equation}\label{eq:Cx}
        \mathbf{y}=\mathbf{C}\mathbf{x},
    \end{equation}
where $\mathbf{C}\in\mathbb{R}^{n\times N}$ is the output matrix whose rank is equal to $n$. For example, should we measure the first generalized coordinate and its time derivative, $\mathbf{C}$ would have two rows: the first with all zeros but a one in the first entry, and the second with all zero but a one in the $(N+1)$th entry.

We consider one experiment, in which the first order system is initialized from $\mathbf{x}_0$. During the experiment, we record $S$ samples of $\mathbf{y}$ from $t=0$ at a finite resolution $\Delta t$. We partition the total recording into consecutive $P$ segments of duration $s=\lfloor S/P \rfloor$. From the segments, we construct the detection matrix $\mathbf{T}\in\mathbb{R}^{ns\times P}$ by juxtaposing measurements for neighboring segments
    \begin{equation}\label{eq:T}
    \mathbf{T}=\left[ 
    \begin{array}{cccc} 
      \tilde{\mathbf{y}}^{(1)} & 
      \tilde{\mathbf{y}}^{(2)} &
      \cdots &
      \tilde{\mathbf{y}}^{(P)}
    \end{array} 
    \right],
    \end{equation}
where, for $p=1,\ldots, P\,$,
    \begin{equation}\label{eq:colT}
    \tilde{\mathbf{y}}^{(p)}=\left[ 
    \begin{array}{c} 
      y_1(s(p-1)\Delta t)\\
      y_1((s(p-1)+1)\Delta t)\\
      \vdots \\
      y_1((sp-1)\Delta t)\\
      \hline\
      \vdots\\
      \hline\
      y_n(s(p-1)\Delta t)\\
      y_n((s(p-1)+1)\Delta t)\\
      \vdots\\
      y_n((sp-1)\Delta t)\\
    \end{array} 
    \right].
    \end{equation}
    
Next, we  {analytically demonstrate} that, under quite general conditions, the rank of the detection matrix converges to $2N$ for a sufficiently long experiment. For convenience, the entries in the generic column in \eqref{eq:colT} can be reordered according to the time stamp, so that all measurements at the same time instant are grouped. This can be done using a simple permutation matrix $\bm{\Pi}\in\{0,1\}^{ns\times ns}$,
    \begin{equation}\label{eq:ytilde}
        \tilde{\mathbf{y}}^{(p)}=\bm{\Pi}
        \left[ 
    \begin{array}{c} 
      \mathbf{y}(s(p-1)\Delta t)\\
      \mathbf{y}((s(p-1)+1)\Delta t)\\
      \vdots \\
       \mathbf{y}((sp-1)\Delta t)
    \end{array} 
    \right].
    \end{equation}
The permutation matrix is nonsingular, since the absolute value of its determinant is always one \cite{bernstein2009matrix}. 

For a linear system, the entire dynamics is captured by the transition matrix $\bm{\Phi}:\mathbb{R}_{\geq 0}\times\mathbb{R}_{\geq 0}\rightarrow \mathbb{R}^{2N\times 2N}$ that maps any initial state at time $\tau$ to a final state at time $t$ \cite{rugh1996linear}.  In the case of a linear time-invariant system, as in \eqref{eq:LTI}, the transition matrix takes the form of an exponential matrix, such that
    \begin{equation}
        \bm{\Phi}(t,\tau)=\exp(\mathbf{A}(t-\tau))
    \end{equation}
Following the same lines of arguments of Ref.~\onlinecite{Porfiri2020}, we can apply Cayley-Hamilton theorem \cite{rugh1996linear} to express the matrix exponential in the form of a linear combination of the first $2N$ powers of $\mathbf{A}$, namely
    \begin{equation}\label{eq:CH}
        \bm{\Phi}(t,\tau)=\sum_{j=0}^{2N-1}\alpha_j(t-\tau)\mathbf{A}^j
    \end{equation}
where $\boldsymbol{\alpha}=[\alpha_0,\ldots,\alpha_{2N-1}]^{\mathrm{T}}$ is a vector of analytic functions that can be written in terms of spectral properties. Specifically, assuming that matrix $\mathbf{A}$ has distinct, complex, eigenvalues $\lambda_1,\ldots,\lambda_{2N}$ and introducing the vector function $\mathbf{E}:\mathbb{R}\rightarrow \mathbb{C}^{2N}$ and the non-singular Vandermonde matrix $\mathbf{V}\in\mathbb{C}^{2N\times 2N}$, we write
    \begin{equation}\label{eq:alphaVE}
        \boldsymbol{\alpha}(t)=\mathbf{V}^{-1}\mathbf{E}(t),
    \end{equation}
with
    \begin{equation}
        \mathbf{E}(t)=\left[
        \begin{array}{ccc} 
      e^{\lambda_1 t} &
      \cdots &
      e^{\lambda_{2N}t}
    \end{array} 
    \right]^{\mathrm{T}},
    \end{equation}
\begin{equation}
\mathbf{V}=
\left[\begin{array}{cccccc} 
    1 & \lambda_1 & \lambda_1^2 & \cdots & \lambda_1^{2N - 2} & \lambda_1^{2N - 1} \\
1 & \lambda_2 & \lambda_2^2 & \cdots & \lambda_2^{2N - 2} & \lambda_2^{2N - 1} \\
\vdots & \vdots & \vdots & \ddots & \vdots & \vdots \\
1 & \lambda_{2N} & \lambda_{2N}^2 & \cdots & \lambda_{2N}^{2N - 2} & \lambda_{2N}^{2N - 1}
\end{array}
\right].
\end{equation}

By utilizing \eqref{eq:CH} for the transition matrix, we express the output at time $t$ in terms of the initial state at $\tau$ as follows:
    \begin{equation}\label{eq:yO}
        \mathbf{y}(t)=
        \sum_{j=0}^{2N-1}\alpha_j(t-\tau)\mathbf{C}\mathbf{A}^j\mathbf{x}(\tau)=
        ( \boldsymbol{\alpha}(t-\tau)\otimes\mathbf{I}_{n\times n})^{\mathrm{T}} \mathcal{O} \mathbf{x}(\tau)
    \end{equation}
where ``$\otimes$'' is the Kronecker product and $\mathcal{O}\in\mathbb{R}^{2nN\times 2N} $ is the observability matrix of the linear system \cite{rugh1996linear}, defined as
    \begin{equation}
         \boldsymbol{\mathcal{O}}=\left[ 
    \begin{array}{c} 
      \mathbf{C}\\
      \mathbf{CA}\\
      \vdots \\
      \mathbf{CA}^{2N-1}
    \end{array} 
    \right].
    \end{equation}

Now, we are ready to express the detection matrix in terms of the key properties of the system, the sampling rate, and the length of the time-series. By inserting \eqref{eq:yO} into \eqref{eq:ytilde} and using \eqref{eq:alphaVE}, we obtain
    \begin{equation}\label{eq:ytilde2}
        \tilde{\mathbf{y}}^{(p)}=\bm{\Pi}(\boldsymbol{E}_{(s)}^{\mathrm{T}}\boldsymbol{V}^{\mathrm{-T}}\otimes\mathbf{I}_{n \times n})\boldsymbol{\mathcal{O}}\mathbf{x}(s(p-1)\Delta t),
    \end{equation}
where $\boldsymbol{E}_{(s)}\in\mathbb{C}^{2N\times s}$ collates the sampled dynamics unfolding in the eigenspaces of the state matrix according to \eqref{eq:alphaVE},
    \begin{equation}
        \boldsymbol{E}_{(s)}=\left[ 
    \begin{array}{cccc} 
      \boldsymbol{E}(0) &
      \boldsymbol{E}(\Delta t) &
      \cdots &
      \boldsymbol{E}((s-1)\Delta t)
    \end{array} 
    \right].
    \end{equation}
Finally, by replacing \eqref{eq:ytilde2} into \eqref{eq:T}, we establish
\begin{equation}\label{eq:T2}
    \mathbf{T}=\bm{\Pi}(\boldsymbol{E}_{(s)}^{\mathrm{T}}\boldsymbol{V}^{\mathrm{-T}}\otimes\mathbf{I}_{n \times n})\boldsymbol{\mathcal{O}}\mathbf{X}_0.
\end{equation}
where $\mathbf{X}_0\in\mathbb{R}^{2N\times P}$ summarizes the initial conditions for all of the segments in which we partitioned the available time-series, that is
    \begin{equation}\label{eq:X0}
        \mathbf{X}_0=\left[ 
    \begin{array}{cccc} 
      \mathbf{x}(0) & 
      \mathbf{x}(s\Delta t) &
      \cdots &
      \mathbf{x}(s(P-1)\Delta t)
    \end{array} 
    \right],
    \end{equation}
    
To  {show} our main claim, we should recall four basic properties of matrix algebra for the computation of the rank of matrices \cite{bernstein2009matrix,gentle2007matrix}. Specifically, given two complex conforming matrices $\mathbf{M}_1$ and $\mathbf{M}_2$, we have
\begin{enumerate}
\item[P1)] $\mathrm{rank}(\mathbf{M}_1\mathbf{M_2})\leq \min\{\mathrm{rank}\mathbf{M_1},\mathrm{rank}\mathbf{M_2}\}$; 

\item[P2)] if $\mathbf{M}_1$ is full column rank, then $\mathrm{rank}(\mathbf{M}_1\mathbf{M_2})=\mathrm{rank}\mathbf{M_2}$; 

\item[P3)] if $\mathbf{M}_1$ is full row rank, then $\mathrm{rank}(\mathbf{M}_2\mathbf{M_1})=\mathrm{rank}\mathbf{M_2}$; 
\item[P4)] $\mathrm{rank}(\mathbf{M}_1\otimes \mathbf{M_2})=\mathrm{rank}\mathbf{M_1}\mathrm{rank}\mathbf{M_2}$.
\end{enumerate}

By virtue of P1 and P2, we can readily bound the rank of the detection matrix as follows:
    \begin{equation}
        \mathrm{rank} \mathbf{T}\leq\mathrm{min}\left\{\mathrm{rank}\left[(\boldsymbol{E}_{(s)}^{\mathrm{T}}\boldsymbol{V}^{\mathrm{-T}}\otimes\mathbf{I}_{n \times n})\boldsymbol{\mathcal{O}}\right],
        \mathrm{rank}\mathbf{X}_0\right\}.
    \end{equation}
The rank of $\mathbf{X}_0$ is less than or equal to any of its dimensions, that is, $2N$ or $P$. {Similarly, the rank of $(\boldsymbol{E}_{(s)}^{\mathrm{T}}\boldsymbol{V}^{\mathrm{-T}}\otimes\mathbf{I}_{n \times n})\boldsymbol{\mathcal{O}}$ must be less or equal than its dimensions, that is, $sn$ or $2N$. As a result the rank of the detection matrix must be less than $2N$}, so that
   \begin{equation}\label{eq:bound}
        N\geq\frac{\mathrm{rank}{\mathbf{T}}}{2}.
    \end{equation}
This inequality ensures that the use of the detection matrix to infer the number of degrees of freedom of a system will always lead a conservative estimate. In other words, no choice of the length of the time-series and number of partitions will overestimate the true number of degrees of freedom that could be predicted from measurement of a partial subset.  

One \textit{caveat} to this last statement is that the estimation of the rank of a matrix is far from being a trivial task, which could, in turn, produce numerical confounds in the estimation of the number of DOFs through the detection matrix. In our work, we estimate the rank from the location of the largest spectral gap, as originally presented in \cite{Haehne2019}. We first examine the ordered singular values $\sigma_j$ of $\mathbf{T}$ (such that $\sigma_1 \geq \sigma_2 \geq \ldots \geq \sigma_j \geq \ldots$) and extract the spectral gap $\Delta_j=\sigma_{j}-\sigma_{j+1}$ for each singular value. Calling $\Delta_{\tilde{N}}$ the maximum gap, the index $\tilde{N}$ will represent an estimate of the rank of $\mathbf{T}$. The singular values are obtained via the \verb+svds+ command in Matlab. A graphical example of such estimation is shown in Fig.~\ref{f:meth}. 

For the rank of the detection matrix to be exactly equal to twice the number of degrees of freedom, we shall formulate  three hypotheses on the quality and quantity of the measurements.
\begin{enumerate}
\item[H1)] $(\mathbf{C},\mathbf{A})$ constitutes an observable pair \cite{rugh1996linear}, so that $\mathrm{rank}\boldsymbol{\mathcal{O}}=2N$. This condition is equivalent to state that none of the eigenvectors of $\mathbf{A}$ are in the null space of $\mathbf{C}$ through the Popov-Belevich-Hautus lemma \cite{rugh1996linear}. For an undamped system or a system with proportional damping, where the damping matrix $\mathbf{D}$ does not alter the natural modes, this condition implies that we shall not take our measurements in correspondence of a node of any of the mode shapes. Some criteria on sensor placement to ensure the observability of  {mechanical systems and engineering structures} are presented in the works of Gawronski and Williams~\cite{Gawronski1991} and Liu \emph{et al.}~\cite{Liu1994}.

\item[H2)] The size of each partition is larger than the dimension of the system, so that $s\geq 2N$. By virtue of this hypothesis and excluding degenerate cases of undamped systems in which the imaginary eigenvalues are spaced by multiples of $2\pi\sqrt{-1}/\Delta t$, $\boldsymbol{E}_{(s)}$ is a non-singular Vandermonde matrix \cite{Porfiri2020} so that $\boldsymbol{E}_{(s)}$ is full row rank, that is,  $\mathrm{rank}\boldsymbol{E}_{(s)}=2N$. By using P2 and recalling that $\mathbf{V}$ is nonsingular, H1 ultimately implies $\mathrm{rank}(\boldsymbol{E}_{(s)}^{\mathrm{T}}\mathbf{V}^{-\mathrm{T}})=2N$.

\item[H3)] The initial conditions for each partition are such that $\mathrm{rank}\mathbf{X}_0=2N$. While it could be difficult to pinpoint at general conditions that ensure the fulfilment of this hypothesis, we offer the following necessary conditions. The first trivial condition is that the number of partitions is larger than the dimension of the system, so that $P\geq 2N$. The second, intuitive assumption, is that the initial condition $\mathbf{x}_0$ should have nonzero projection on all the eigenspaces of the state matrix; for example, should we pick an initial condition on one of the eigenspaces, we would obtain that all the columns are linearly dependent. In the case of undamped systems or systems with proportional damping, we shall exclude initial conditions that that would not excite all the natural modes of the system. Stating broader conditions would require the study of a generalized Vandermonde matrix which might be rather case specific. 
\end{enumerate}

Under H1--H3, application of P2--P4 to \eqref{eq:T2} ensures that the inequality in \eqref{eq:bound} becomes an equality.
Hence, from time-series of a single experiment it is possible to exactly infer the total number of degrees of freedom of the entire system. In practical terms, all of these hypotheses are quite mild, whereby the set of unobservable pairs has measure equal to zero such that H1 would be always satisfied when working with experimental data \cite{sontag2013mathematical}. Likewise, working with experimental time-series, it is tenable that a random initial condition will excite the dynamics of the totality of the eigenspaces and that eigenvalues of the state matrix will not have a perfect equal spacing on the imaginary axis. Hence, by acquiring sufficiently long time-series and varying the number of partitions, one should expect a close estimate of the number of degrees of freedom.

The use of the detection matrix requires neither the presence nor the knowledge of any input to the system. We only need access to one transient response, rather than the steady-state dynamics elicited by stationary input signals that are at the core of many established SHM approaches~\cite{Doebling1998}. Should one have access to some control input, it would be possible to contemplate alternative techniques such as the one recently proposed by Tyloo and Delabays~\cite{Tyloo2021},
which relies on a single harmonic input for successfully undertaking the problem of network size inference in the case of diffusive coupling -- with respect to a  {mechanical system or an engineering structure}, this would result in the need for a rigid body motion.

An equivalent  {derivation} can be carried out by releasing the assumption of time-invariance, following the arguments in the Supplemental Material of Ref.~\onlinecite{Porfiri2020}. The main argument of the proof would be based on the extension of the notion of observability matrix to analytic systems with constant output matrix, which allows one to build a representation of the transition matrix similar to \eqref{eq:CH}.

\subsection{Detection matrix vs. Hankel matrix}
\label{s:equivalence}

As mentioned in the introduction, the detection matrix has origins in network theory. However, this matrix is similar in nature to  block Hankel matrices used in the study of discrete-time dynamical systems, such as those underlying 
dynamic mode decomposition in fluid flows~\cite{tu2014dynamic} and the inference of network size from partial measurements~\cite{tang2020dynamical}. Hankel matrices for discrete-time systems are also at the core of subspace identification theory~\cite{van2012subspace}. Output-only versions of such matrices have been used extensively in damage detection, and they form the basis of stochastic subspace identification techniques~\cite{Peeters1999, Deraemaeker2008}. 

In Appendix~\ref{a:equivalence}, we demonstrate that the block Hankel matrix assembled from the transient dynamics of a discrete-time system in the absence of any input can be viewed as a detection matrix. The network-theoretic concept of detection matrix applies to a wide class of continuous-time systems, while the Hankel matrix in subspace identification theory is typically defined for discrete-time systems that arise from Euler-type discretizations. For these discretizations to be valid, the time-step must be sufficiently small with respect to the time-scale of the system. In contrast, no specification on the time resolution is needed for the theoretical development of the detection matrix and its use even extends to time-varying dynamics, where the underlying properties of the system vary in time.

\begin{figure*}[!htb]
\centering
\includegraphics[scale=1.1]{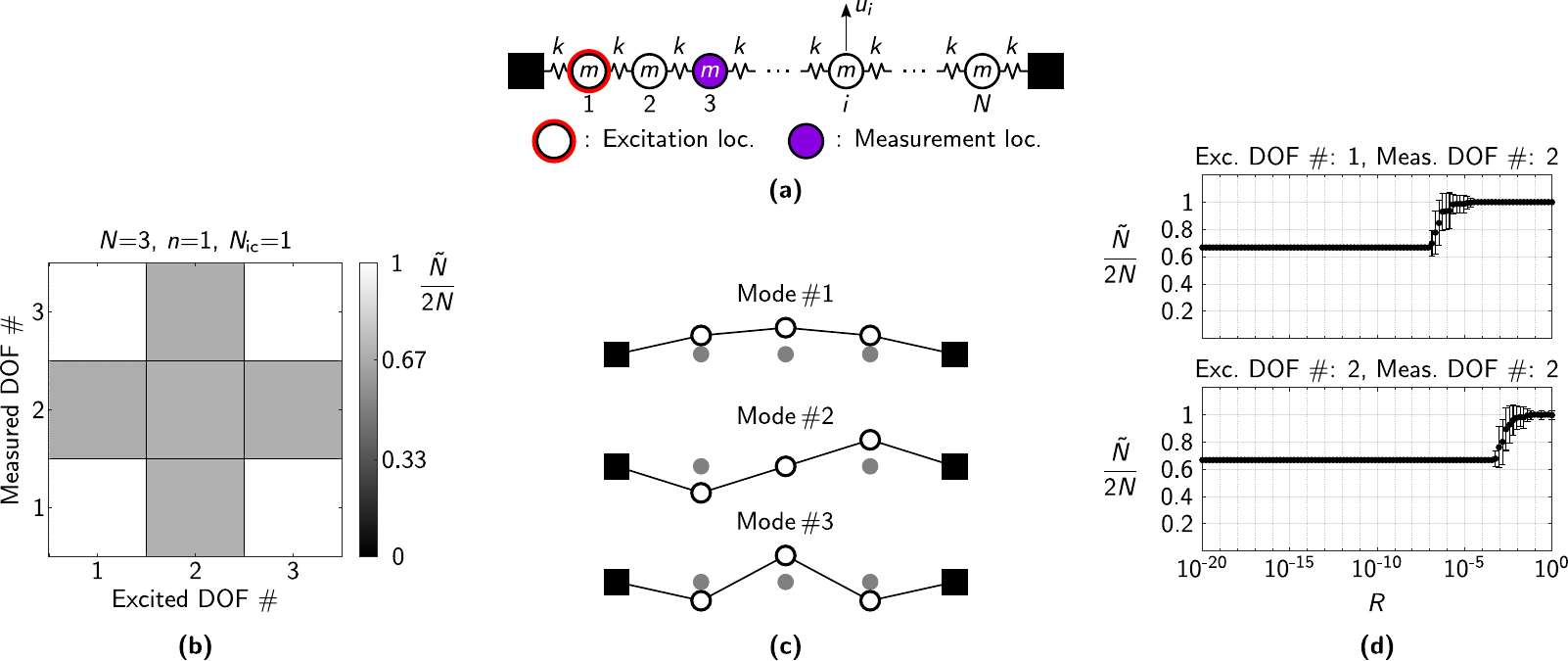}
\caption{(a) Schematic of the spring-mass lattice system with one DOF per mass. (b) Dependency of the {normalized size estimate $\tilde{N}/(2N)$} on the number of excited and measured dofs, for $N=3$. The system is excited with an initial displacement at a single location, and we measure the displacement time trace at a single location. (c) Mode shapes for the system in (b), showing that the second mass is located at a node of mode \#2. (d) Effects of stiffness randomization on {$\tilde{N}/(2N)$}, for two combinations of excitation and measurement locations. Each point is the average of 100 realizations and the error bars indicate the standard deviation.}
\label{f:chain1}
\end{figure*}

\section{Application to spring-mass lattices}
\label{s:dem}

We first discuss the application of the detection matrix to 1D lattices of springs and masses. This simple problem allows us to assess the robustness of the inferences made through $\mathbf{T}$ with respect to measurement and acquisition-related parameters as well as physical phenomena that could strain the applicability of the method, while avoiding issues related to the complexity of the system topology. In addition, we show how the detection matrix can be used to locate weakened springs in structural lattices of known size -- the most preliminary step towards the application of this technique to SHM.

\subsection{Details on the system and the simulation method}
\label{s:sim}

We consider a 1D lattice of $N$ masses connected by $N+1$ springs (Fig.~\ref{f:chain1}a), in which adjacent masses are connected by springs. We use the displacements of each mass, $u_i$,  $i=1,2,\,...\,N$ as the generalized coordinates. The first and last masses of the lattice are connected to ground; unless otherwise specified, all masses and springs are identical ($m=0.1\,\mathrm{kg}$ and $k=1\,\mathrm{Nm^{-1}}$). Unless otherwise specified, we set values of damping to zero; when we consider damping, we assume that this is of the Rayleigh type and proportional to the mass and stiffness matrices, so that,
    \begin{equation}
    \mathbf{D}=d_1\mathbf{M}+d_2\mathbf{K}
    \end{equation}
where $d_1,d_2\in\mathbb{R}_{\geq 0}$ are proportional damping coefficients that ensure the damping matrix does not alter the natural modes of the system.

Throughout Section~\ref{s:dem}, we only apply displacements as initial conditions and only record displacements. However, similar results can be obtained from velocities or from a combination of velocities and displacements. The initial condition for the lattice, $\mathbf{x}_0=\left[\mathbf{u}_0^\mathrm{T}, \mathbf{\dot{u}}_0^\mathrm{T}\right]^\mathrm{T}$, is such that all the velocities are equal to zero and only a subset of  $N_\mathrm{ic}$ generalized coordinates are excited.  {The initial value for the displacement is set within the range $[-U,U]$, where $U=1\,\mathrm{m}$; such a choice bears no consequences on the entire linear analysis.} We refer to the masses whose initial displacement is nonzero as the ``excitation locations'' in the figure. The  $n$ masses whose displacement is being recorded are called the ``measurement locations''. The number of excited DOFs is $N_\mathrm{ic}$, while the number of measurement locations is coincident with the number of recorded time traces $n$. 

Synthetic data upon which we test the validity of the detection matrix are obtained through modal analysis, to avoid introducing errors due to time integration. First, we determine the mass and stiffness matrices of the system, $\textbf{M}$ and $\textbf{K}$, and calculate the natural radian frequencies $\omega_i\in\mathbb{R}_{>0}$ (assumed to be different from each other) and natural modes $\boldsymbol{\psi}_i\in\mathbb{R}^N$, where $i=1,2,\,...\,N$. Then, we use the following closed-form expression for the time evolution of the displacement~\cite{Meirovitch2010}:
\begin{multline}
    \textbf{u}(t)=\sum_{i=1}^{N} \boldsymbol{\psi}_i\,e^{-\xi_i \omega_i t}\left[ \frac{\boldsymbol{\psi}_i^{\mathrm{T}}\,\textbf{M}\,\textbf{u}_0}{\mu_i} \cos{(\omega_{di} t)} \right. \\
    \left. + \frac{\boldsymbol{\psi}_i^{\mathrm{T}}\,\textbf{M}}{\mu_i\,\omega_{di}} \left( \dot{\textbf{u}}_0 + \xi_i \, \omega_i \, \textbf{u}_0 \right)\,\sin{(\omega_{di} t)} \right]\,,
\end{multline}
where $\mu_i={\psi}_i^{\mathrm{T}} \mathbf{M}{\psi}_i$ are the modal masses, $\xi_i=d_1/(2\omega_i)+d_2\omega_i/2$ are the modal damping ratios, and $\omega_{di}=\omega_i\sqrt{1-\xi_i^2}$ are the natural frequencies of damped vibrations. In the simulations, the proportional damping coefficients $d_1$ and $d_2$ are chosen so that the values of $\xi_i$ associated with the first two modes, $\xi_1$ and $\xi_2$, are equal to each other and to a prescribed value.

The excitation locations are predetermined, and the values of initial displacements are randomly generated within the chosen interval. Simulations are carried out for a total of $S$ time instants, and the time-step is set to $\Delta t=T_0/(10\,\pi)$, where $T_0=2\pi/\omega_0$ and $\omega_0=\sqrt{k/m}$. In general, our time acquisition parameters are chosen to respect hypotheses H2 and H3 in Section~\ref{s:theory}.

In the following, we consider two lattice configurations: 1) a small system with $N=3$ DOFs, characterized by (normalized) natural radian frequencies $1/\omega_0\{ 0.765$,$1.414$,$1.848 \}$, and 2) a large system with $N=25$, featuring natural frequencies $1/\omega_0\{ 0.121,0.241,0.361,0.479,0.595,0.709,0.821,0.929,\allowbreak 1.035,\ldots,1.996\}$.

\subsection{Importance of excitation and measurement locations: illustration on a 3-DOFs system}
\label{s:1Dobsl}

To investigate how the excitation and measurement locations affect {$\tilde{N}$, the location of the maximum spectral gap}, we consider a small system of only three masses, in which we excite a single DOF and record the time trace of a single DOF ($N=3$ and $N_\mathrm{ic}=n=1$). We let the simulation run for  approximately 19 times the characteristic period of the system ($S \Delta t/T_0\approx 19$), and choose $P=9$ signal partitions. The normalized {size of the system, $\tilde{N}/(2N)$}, as a function of the choice of excited and measured DOFs is reported as a color map in Fig.~\ref{f:chain1}b; each square of the colormap is the averaged result of 100 simulations with randomly-varying initial displacement values. 

One can see that the normalized estimate is equal to one (the method allows to perfectly infer the number of masses in the system) only if neither the measured nor the excited DOFs corresponds to the second mass, the central mass of the system. This can be explained by looking at the mode shapes shown in Fig.~\ref{f:chain1}c, which challenge the validity of H1 on the observability of $(\mathbf{A},\mathbf{C})$ or of H3 on the rank of $\mathbf{X}_0$. 

Clearly, the second mass is a zero-displacement location or node of this perfectly-symmetric system. In particular, should we select the second mass as the measurement location, we would obtain the following observability matrix in SI units:
    \begin{equation}
        \boldsymbol{\mathcal{O}}= \left[ \begin{array}{cccccc} 
        0  & 1    &    0     &   0     &   0     &    0\\
         0     &    0   &      0    &     0  &  1      &   0\\
    1/2 &  -1 &   1/2    &     0     &    0  &      0\\
         0  &       0   &      0 &    1/2 &   -1  &  1/2\\
   -2/3 &   1 &  -2/3   &      0    &     0     &    0\\
         0      &   0   &      0 &  -2/3 &    1 &   -2/3
        \end{array}\right].
    \end{equation}
Such a matrix has rank equal to four, thereby hindering the successful inference of the exact number of DOFs of the lattice. 

Likewise, exciting the mass at the center will systematically lead to the entries in the matrix $\mathbf{X}_0$ corresponding to the first mass to be equal to those corresponding to the third mass, due to the symmetry of the system. As a result, the rank of $\mathbf{X}_0$ is only four and our method cannot achieve perfect inference of the system size. 

These considerations demonstrate that the excitation and measurement locations are of critical importance for the successful identification of the size of the system, at least when working with toy examples that could contain perfect symmetries or the emergence of nodes.

\subsection{Beneficial role of imperfections: illustration on a 3-DOFs system}
\label{s:1Dloc}

Real  {systems} are never perfectly symmetric due to their fabrication and assembly processes. These unavoidable imperfections are actually beneficial to the success of our method, as they prevent the observability matrix $\boldsymbol{\mathcal{O}}$ and the matrix of initial conditions $\mathbf{X}_0$ to be rank deficient. In the following, we carry out this argument for the same 3-DOF system discussed above. In particular, we choose to separately perturb the stiffness of each spring by multiplying it by $(1+r)$, where $r$ is a uniform random variable between 0 and $R$. For each value of $R$, we perform 100 simulations and record the average and standard deviation of {$\tilde{N}/(2N)$}; we repeat this process for two combinations of excited and measured DOFs and report the results in Fig.~\ref{f:chain1}d.

In the top panel of Fig.~\ref{f:chain1}d, we consider the case where we excite the first mass and record the time trace of the second mass, thereby violating H1 in the absence of defects. As discussed in Section~\ref{s:1Dobsl}, this choice does not yield perfect inference for the defect-free system. For {$R \approx 10^{\text{-}6}$}, we observe a transition between imperfect and perfect inference. This result has significant implications from an application perspective, and it indicates that issues related to the selection of excited and measured DOFs are mitigated by the presence of extremely small imperfections.  This transition occurs since the second mass can no longer be a node, as the system loses symmetry due to randomizations of its properties.

We also consider the extreme case where the second mass is chosen for both excitation and measurement, thereby violating both H1 and H3 in the absence of imperfections. As shown in Fig.~\ref{f:chain1}d, very small defects are sufficient to accurately infer the exact number of DOFs in the system. However, a larger value of $R \approx 10^{\text{-}2}$ would be needed in comparison to the previous case.

\subsection{Role of the number of measured time traces in longer lattices}
\label{s:1Dobsn}

Our intuition tells us that the number of measured time traces should play an important role on the effectiveness of the method, especially for systems with many DOFs. While a single time trace might suffice to infer the size of a system with a few DOFs (such as the 3-DOFs system above), the same might not be true for more complex systems. Surprisingly, this influence is not as dramatic as one might expect, and, in practice, the inference problem can be successfully addressed through a small fraction of measurement locations.

\begin{figure*}[!htb]
\centering
\includegraphics[scale=1.1]{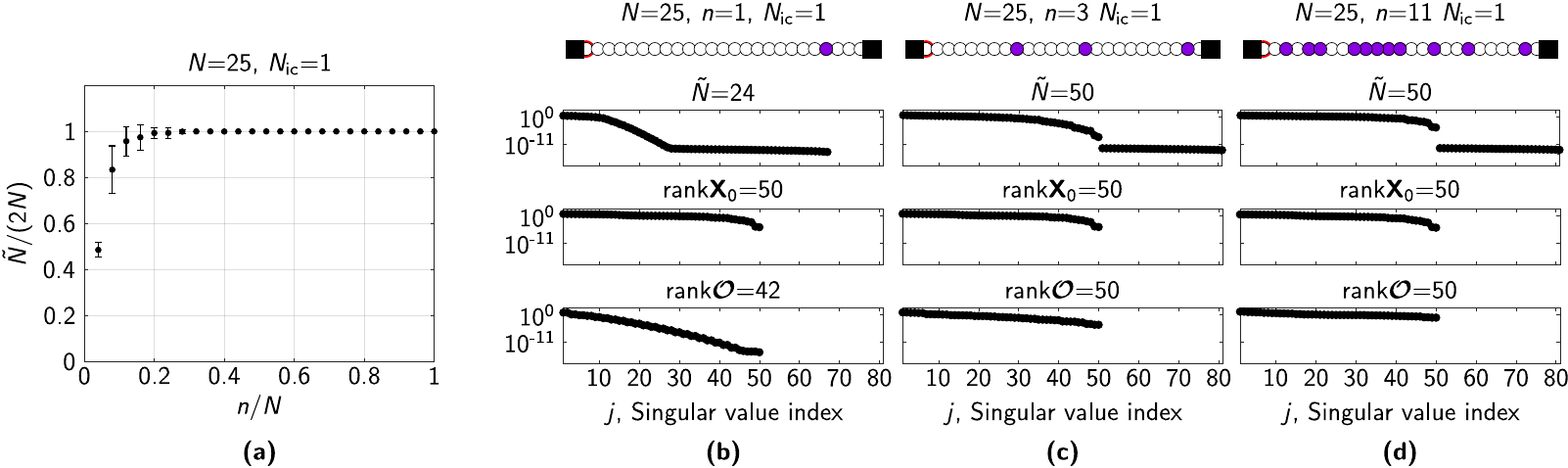}
\caption{Effect of the number of measured time traces on a lattice with $N=25$ and $N_\mathrm{ic}=1$. In all subplots, we excite the system at the leftmost mass with an initial displacement, and we only record displacements; moreover, we set $S \Delta t/T_0=175$ and $P=81$. (a) Evolution of the normalized {size estimate} as a function of the number of recorded time traces, $n$. Each point is the average of 100 realizations and the error bars indicate the standard deviation. (b-d) Singular value portraits of {(from top to bottom)} the detection matrix $\mathbf{T}$, the initial conditions matrix $\mathbf{X}_0$ and the observation matrix $\boldsymbol{\mathcal{O}}$ for different $n$ values. For the case at hand, in which $P\geq2N$ and $n\geq1$, matrices $\mathbf{X}_0$ have $2N$ singular values; on the other hand $\mathbf{T}$ has 67 singular values for $n=1$ and 81 for $n\geq 2$.}
\label{f:chain2}
\end{figure*}

We consider a 25-mass-long lattice ($N=25$), and we apply an initial displacement to the very first mass near the left clamp, such that $N_\mathrm{ic}=1$. We make this choice since the motion of the corresponding mass is nonzero for all natural modes of the system, thereby favoring the validity of H3 ($\mathrm{rank}\mathbf{X}_0=2N$). We then let $n$ vary between 1 and $N$, and we record {$\tilde{N}$, the location of the maximum spectral gap}. While the total available number of time traces is $2N$ since we can record both displacements and velocities, we here opt to only record displacements. For each value of $n$, we perform 100 simulations, where we change the value of the initial condition and the measurement locations in a randomized fashion, and we record the average and standard deviation of {$\tilde{N}/(2N)$}. In each simulation, we select a number of time samples such that $S \Delta t/T_0=175$, and $P=81$ partitions. The results of this procedure are shown in Fig.~\ref{f:chain2}a.
We observe that choosing $n=1$ (corresponding to $n/N=0.04$) yields a large underestimation of the size of the system that is predicted to be around 12, rather than 25. Increasing $n$ to 2 allows to achieve more accurate inferences -- albeit with some modest variability that depends on the measurement location. As expected, increasing $n$ further will beget perfect inference with a negligible standard deviation. 

To gain a better understanding on the cause of discrepancies from perfect inference, we examine the rank of both matrices $\mathbf{X}_0$ and $\boldsymbol{\mathcal{O}}$, which underlie the validity of H1 and H3. For completeness, we also plot their singular values as obtained from Matlab.  We first consider the case where $n=1$, keeping the same conditions used to obtain Fig.~\ref{f:chain2}a; this result is shown in Fig.~\ref{f:chain2}b. We register that $\boldsymbol{\mathcal{O}}$ is rank deficient, thereby contradicting H1. Thus, not surprisingly, {$\tilde{N}$} does not converge to $2N$.  

If we increase $n$ to 3, as shown in Fig.~\ref{f:chain2}c, $\boldsymbol{\mathcal{O}}$ becomes fully ranked and $\mathbf{T}$ follows suit. {We observe that a distinct spectral gap appears in $\mathbf{T}$; this gap increases as we further increase the number of measured time traces to 11, as shown in Fig.~\ref{f:chain2}d}. From these plots we can observe that, while $\tilde{N}$ and other rank estimates provide some ``monochromatic'' information on the size of the system, the singular value portrait of the detection matrix provides further information on the quality of our inference. As discussed in Section~\ref{s:weak}, the richness of information carried by this portrait can be leveraged for preliminary damage detection.

\subsection{Influence of signal acquisition parameters and damping}

\begin{figure*}[!htb]
\centering
\includegraphics[scale=1.1]{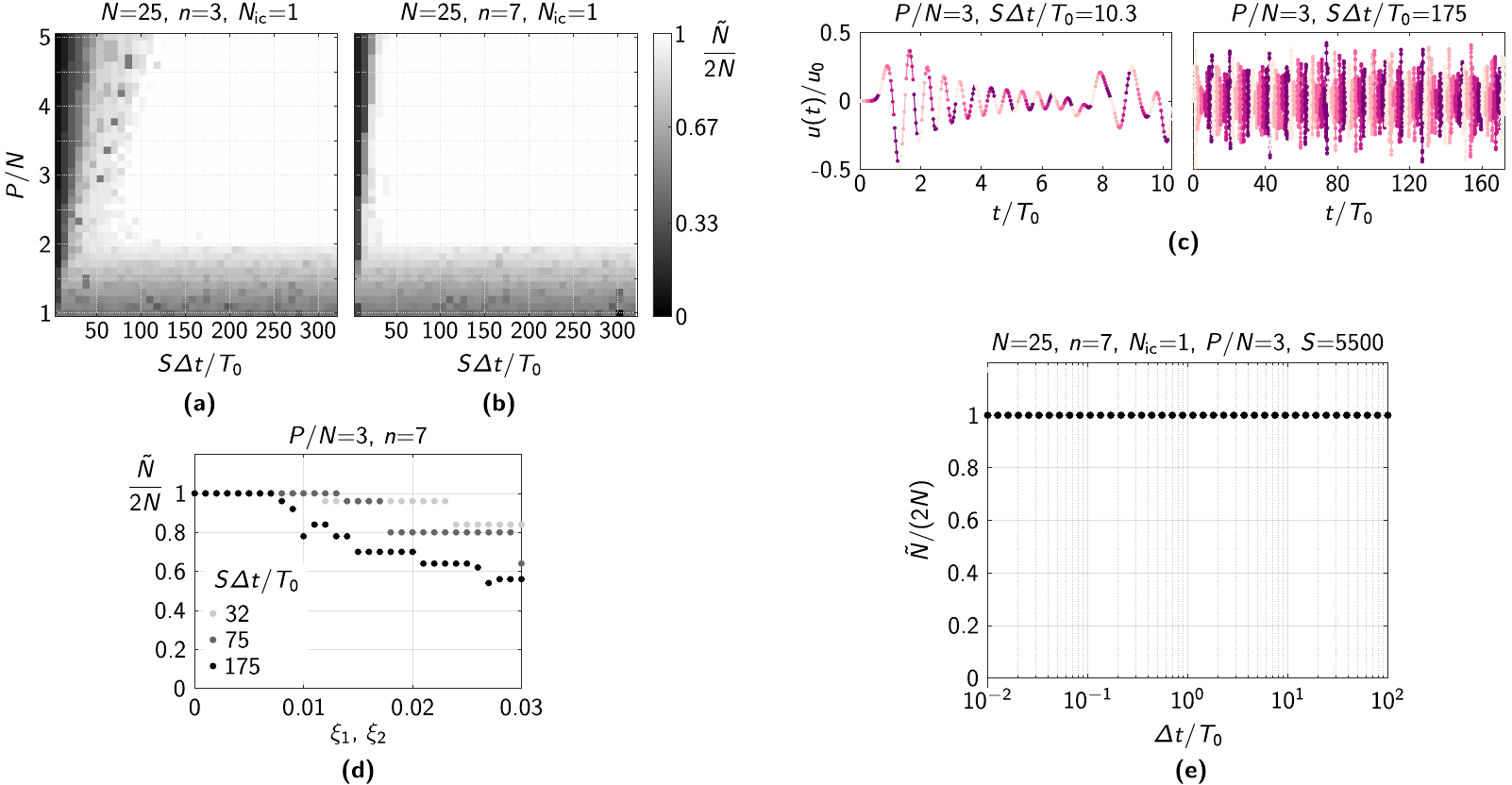}
\caption{Effects of the signal acquisition parameters on a lattice with $N=25$ and $N_\mathrm{ic}=1$. In all subplots, we excite the system at the leftmost node with an initial displacement and we only record displacements. (a,b) Colormaps representing the normalized size estimate {$\tilde{N}$} as a function of the total length of the acquired signal $S$ and the number of partitions $P$. $\Delta t$ is kept fixed. Each point is the average of 10 realizations. (c) Segmented time-series recorded at one of the observation nodes for two sets of acquisition parameters (normalized by the initial condition of the excitation node). Each segment is entered in a separate column of $\mathbf{T}$.  (d) Influence of damping on {$\tilde{N}/(2N)$} for various signal lengths. Each point is the average of 100 realizations. {(e) Dependency of the normalized estimate $\tilde{N}/(2N)$ on the time-step $\Delta t$, fixing $S=5500$ and $P/N=3$}.}
\label{f:chain3}
\end{figure*}

Here, we study the influence of the time acquisition parameters. To do so, we consider the same lattice with $N=25$ and $N_\mathrm{ic}=1$. We then fix $n$ and perform simulations for various nondimensional signal lengths $S\Delta t/T_0$ and numbers of partitions $P$. For each value of the two parameters, we perform 10 simulations with randomized initial condition values and measurement locations, and we record the average {of the normalized estimate $\tilde{N}/(2N)$}. These averages are then plotted as colormaps in Fig.~\ref{f:chain3}a,b for $n=3$ and $n=7$, respectively.

In all these simulations, we consider no damping. Both figures clearly show regions where the detection matrix underestimates the size of the system. One of these regions is the band corresponding to $P<2N$, a condition that violates H3. In addition, the detection matrix does not yield perfect inference for ranges of parameters corresponding to the dark triangles on the left sides of Fig.~\ref{f:chain3}a,b. To understand the origin of these regions, we plot the recorded time-series for two sets of acquisition parameters in Fig.~\ref{f:chain3}c. The left panel corresponds to a point inside the under-inferred triangular region, while the right panel corresponds to a case that yields perfect inference. We can clearly tell that, in the left panel, the signals in each partition are too similar with respect to each other; in turn, the rank algorithm fails to identify the partitioned signals as linearly independent. On the other hand, the partitions of the signal in the right panel feature several oscillations about equilibrium, and are more easily distinguishable by the rank algorithm. From these observations, one could conclude that the total signal length should be chosen to be as large as possible when $N$ is unknown.

The idea of choosing $S$ to be as large as possible fails in the presence of damping. In fact, damping causes the amplitude of the measured signal to decay to zero. This can negatively affect the rank operation since all partitions in the zero-region of the signal are linearly dependent. The influence of damping is spelled out in Fig.~\ref{f:chain3}d, where we plot {$\tilde{N}/(2N)$} as a function of the damping factors of the first two modes ($\xi_1=\xi_2$), for various signal lengths. We can clearly see that the shorter the signal (that is, the smaller $S\Delta t/T_0$), the larger the range of damping values that still yield perfect inference. For example, for {  short and intermediate-length time-series ($S\Delta t/T_0=32$ and 75, respectively)}, perfect inference is achieved for damping factors up to { 0.011}, compared to the maximum damping factor of 0.007 for {  much} longer signals ($S\Delta t/T_0=175$). 

In realistic situations, there are no easy ways of knowing \textit{a priori} the acquisition parameters that work for a  {system} of unknown size. Thus, a user should always inspect the measured signals before trusting the outcome of our method to make sure that damping is not playing a heavy role. In principle, one could still choose to record a very long time-series, but could then discard the part of the signal that is strongly affected by damping. The fact that from theory it is impossible to overestimate the number of DOFs of the system using the detection matrix, as shown in \eqref{eq:bound}, could be used to the user's advantage whereby they could systematically explore different parameter ranges and ultimately consider the ones that yield the largest estimate. 

Finally, we consider the influence of the time-step $\Delta t$ for the case of undamped dynamics. In contrast with the typical use of block Hankel matrices that call for small time-steps to discretize the system dynamics using an Euler scheme, the detection matrix is based on a continuous-time representation. From a theoretical perspective, the consideration of a full, non-discretized, model justifies the application of the detection matrix for arbitrary time resolutions at which the output is sampled. As shown in Fig.~\ref{f:chain3}e, we confirm the time-step has no effect on the estimate of the number of degrees of freedom (keeping everything else, including the number of samples $S$, constant).  We have no indication of the existence of a Nyquist-like condition for the validity of the detection matrix.  The reasons are likely in our use of rich, transient dynamics and in the lack of a need to reconstruct the entire system dynamics. For example, for $\Delta t/T_0=1$, there are only four  {natural} modes with natural frequencies below one half of the sampling frequency (the limit set by Nyquist condition); yet, we accurately recover the exact number of DOFs. Thus, exact inference is attained for a wide range of time-steps, including slow sampling times for which Euler discretization may fail.

\subsection{Influence of signal noise}
\label{s:noise}

Another important aspect related to the applicability of the detection matrix in real-world scenarios is the presence of noise, which is bound to make all columns of the detection matrix linearly independent. To investigate the effects of noise on the detection matrix, we consider again the lattice with $N=25$, excited at its leftmost node with an initial displacement. This time, we fix $n=7$ measurement locations. A white Gaussian noise is added to the time traces recorded in our simulations; this is generated in Matlab by means of the \verb+wgn+ function. The power of the noise added to each time trace, in decibel (dB), is calculated as $P_\mathrm{noise, dB}=P_\mathrm{signal, dB}-\mathrm{SNR_{dB}}$, where $P_\mathrm{signal, dB}$ is the power of each time trace in dB and $\mathrm{SNR_{dB}}$ is a prescribed signal-to-noise ratio. 
\begin{figure}[!htb]
\centering
\includegraphics[scale=1.1]{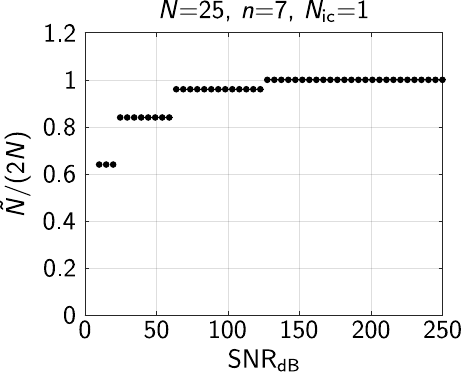}
\caption{{Effect of signal noise on the estimated size of the system. All simulations are performed on the same lattice with $N=25$, $n=7$, and $N_{\mathrm{ic}}=1$. Note that we excite the system at the leftmost node with an initial displacement and we only record displacements. A white noise with a specific signal to noise ratio in decibel, $\mathrm{SNR_{dB}}$, is then added to the signals.}}
\label{f:chainn}
\end{figure}

The dependency of $\tilde{N}/(2N)$ on $\mathrm{SNR_{dB}}$ is illustrated in Fig.~\ref{f:chainn}. Predictably, for sufficiently large values of  $\mathrm{SNR_{dB}}$, above 120\,dB, the method yields perfect inference. Intuition may suggest that increasing noise could result in overestimating the number of degrees of freedom, by contributing spurious dynamics that the method may misinterpret as high-frequency modes. However, this is not the case: increasing signal noise (decreasing $\mathrm{SNR_{dB}}$) causes an underestimation of the number of DOFS with a robust monotonic dependence. Likely, this is due inference of the rank from the spectral gap, which helps mitigate confounding effects due to added noise.

\subsection{Preliminary results on health monitoring and the effects of spring weakening}
\label{s:weak}

\begin{figure*}[!htb]
\centering
\includegraphics[scale=1.1]{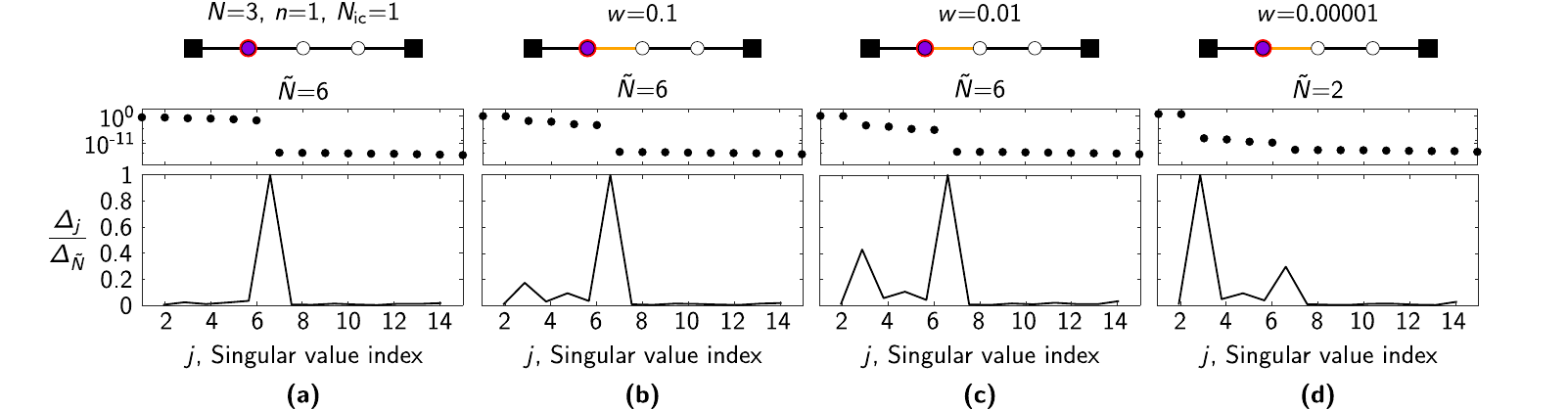}
\caption{Signatures of spring weakening on the singular values of the detection matrix. All simulations are performed on a lattice with $N=3$, $n=1$, and $N_{\mathrm{ic}}=1$, for various degree of weakening of the second spring from the left (highlighted in yellow). {Each subfigure, from top to bottom, shows the considered  {system}, the singular value portrait of $\mathbf{T}$, and the normalized spectral gaps.} (a) Pristine lattice. (b,c,d) Lattices where the stiffness of the above-mentioned spring is $0.1\,k$, $0.01\,k$, and $0.00001\,k$, respectively.}
\label{f:chainw}
\end{figure*}

At this stage, we have all the tools to present some preliminary results on the applicability of the detection matrix for SHM purposes. To do so, we consider the small system with $N=3$. We apply an initial displacement to leftmost mass and record the displacement time-history at the same location. This location is chosen so that our method yields perfect inference for a pristine  {system}, as shown in Fig.~\ref{f:chainw}a, where we report the singular value portrait of $\mathbf{T}$. Below the singular values, we also plot the normalized spectral gaps $\Delta_j/\Delta_{\tilde{N}}$, where $\Delta_{\tilde{N}}$ is the maximum gap. The pristine  {system} has one clear maximum gap corresponding to $\tilde{N}=6$, as illustrated by the single spike in the latter plot. In this undamaged state, we have perfect inference.

Then, we weaken the spring connecting the first and second masses by multiplying its stiffness by a value $w\in (0,1)$. Fig.~\ref{f:chainw}b--d illustrate how weakening values of 0.1, 0.01, and 0.00001 affect singular values and their spectral gaps. In particular, we observe that weakening the spring (decreasing $w$) causes the appearance of a second spike in the spectral gap plot, at $j=2$. As the spring becomes very weak, for $w=0.00001$, we register a clear signature of damage and the spike at $j=2$ becomes the dominant one. Thus, our algorithm outputs $\tilde{N}=2$ and underestimates the size of the system.
For simple  {systems}, like our 1D lattice, this information can point us towards the weakened link. In our example, the fact that two singular values stand out from the others indicates that we are detecting a single mass; indeed, weakening the second link from the left isolates the mass we use for both excitation and measurement, causing us to detect two available DOFs. In complex  {engineering} structures, the damage location may be triangulated with multiple experiments. More detailed studies on damage detection, which would require the development of \textit{ad hoc} damage detection algorithms, deserve a separate treatment that will be part of future studies.

\subsection{Effects of cyclic property variations}
\label{s:cycle}

One of the issues of damage detection algorithms is their susceptibility to changes in the properties of the system of interest. For example, changes in temperature can lead to changes of the stiffness. To handle them, subspace identification methods might require thermal models to infer temperature-adjusted properties~\cite{Basseville2010}; this, however, requires knowledge of the system at hand. It is therefore of interest to evaluate the performance of our model-agnostic estimation tool in the presence of periodic variations of the systems' parameters. To do so, we modify \eqref{eq:eqmot} to account for a time-varying uniform modulation of the stiffness matrix
\begin{equation}\label{eq:eqmotvar}
    \mathbf{M}\ddot{\mathbf{u}}(t)+\mathbf{D}\dot{\mathbf{u}}(t)+\left(1+\gamma \sin{\left( \omega_m t + \phi_m\right)}\right)\mathbf{K}\mathbf{u}(t)=\mathbf{0}_N,
\end{equation}
where $\gamma$ is a modulation amplitude factor, $\omega_m$ is the modulation frequency and $\phi_m$ is the phase of the modulation. This is an example of Mathieu's equation~\cite{Kovacic2018}, and modal analysis techniques cannot be used to derive an analytical solution to it. Thus, its response to initial conditions is found numerically using the \verb+ode45+ function in Matlab.

\begin{figure}[!htb]
\centering
\includegraphics[scale=1.1]{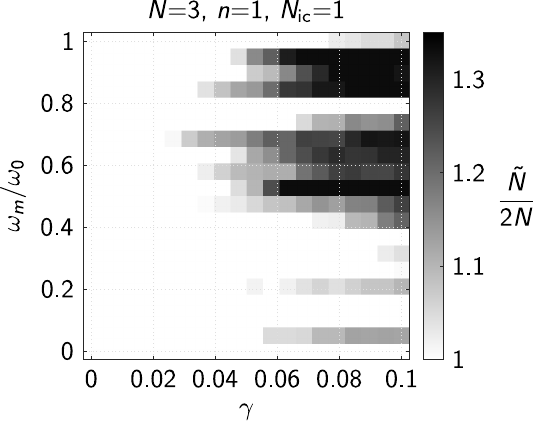}
\caption{{Effect of the modulation parameters on the estimated size of a system with time-varying stiffness. All simulations are performed on a lattice with $N=3$, $n=1$, and $N_{\mathrm{ic}}=1$. $\omega_m$ is the modulation frequency and $\gamma$ is the amplitude of modulation, as shown in \eqref{eq:eqmotvar}. Each point of the colormap is the average of 100 realizations obtained with randomized amplitude of the initial condition and randomized value of the modulation phase $\phi_m$.}}
\label{f:chainmod}
\end{figure}

As a representative example, we again consider the small three-mass lattice, where we excite and record the response at the leftmost node. We consider various values of the modulation parameters $\gamma$ and $\omega_m$; for each pair of values, we solve \eqref{eq:eqmotvar} 100 times, compile the detection matrix, and record the average of $\tilde{N}/(2N)$. For each realization, we randomize the initial condition amplitude and modulation phase $\phi_m$ (we let $\phi_m$ vary to show that our results are independent on the position of the measurement time interval with respect to the property-modulating cycle). The results of this analysis are reported in Fig.~\ref{f:chainmod}.

For low $\gamma$ and $\omega_m$, that is, for small changes of the properties and slow modulation time-scales, our estimation is not affected by time-variations in the stiffness -- a promising evidence for damage detection applications, where property variations for  {systems} of practical relevance unfold along slow time-scales. At large modulation amplitudes and faster time-scales, which are of less practical interest, we identify regimes where the number of degrees of freedom is overestimated by our algorithm. This is not surprising, since Mathieu's equation is known to display instabilities as the modulation parameters increase in amplitude~\cite{Kovacic2018}. 

{ 
\subsection{Effects of nonlinearities}
\label{s:nl}

To conclude our simulation campaign on the 1D spring-mass lattice, we examine the effect of cubic nonlinearities on the validity of the inference through the detection matrix. Cubic nonlineariries are routinely used to describe stiffening and softening effects in mechanical systems and engineering structures, thereby leading to a representation in the form of an unforced Duffing system~\cite{nayfeh2008applied,harne2017harnessing}. In the presence of these nonlinearities, the governing equations  become
\begin{equation}\label{eq:eqmotvar}
    \mathbf{M}\ddot{\mathbf{u}}(t)+\mathbf{D}\dot{\mathbf{u}}(t)+\overline{\mathbf{K}}(\mathbf{u}(t))\mathbf{u}(t)=\mathbf{0}_N,
\end{equation}
where the matrix $\overline{\mathbf{K}}(\mathbf{u}(t))$ replaces the constant matrix $\mathbf{K}$ so that the force exerted by the spring between mass $i$ and $i+1$ will change from $k (u_{i+1}(t)-u_i(t))$ to $k (u_{i+1}(t)-u_i(t))+\overline{k}(u_{i+1}(t)-u_i(t))^3$, $\overline{k}$ being the extent of the nonlinearity.

For simplicity, we focus again on the small system with $N=3$, exciting and recording the response at the leftmost node. Simulations are conducted for different values of $\overline{k}$, so that $\overline{k} U^2/k$ ranges from $-0.5$ to $0.5$ to capture softening and stiffening effects at varying magnitudes. Results in Fig.~\ref{f:chainnl} demonstrate that the detection matrix is successful in the exact inference of the number of DOFs of the system for a wide range of nonlinearities, spanning stiffening (positive $\overline{k}$) and softening (negative $\overline{k}$) scenarios. 
\begin{figure}[!htb]
\centering
\includegraphics[scale=1.1]{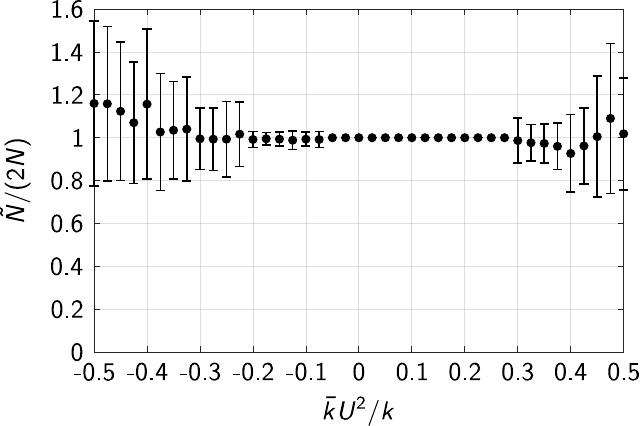}
\caption{{ Effect of cubic nonlinearity on the estimated size of the system. All simulations are performed on a lattice with $N=3$, $n=1$, and $N_{\mathrm{ic}}=1$. $\overline{k} U^2/k$ is the normalized extent of the nonlinearity, defined in Section~\ref{s:nl}; negative and positive values yield softening and hardening nonlinearities, respectively. Each point is the average of 100 realizations obtained with randomized amplitude of the initial condition, and the error bars indicate the standard deviation.}}
\label{f:chainnl}
\end{figure}
As the extent of the nonlinearities approaches $-0.2$ and $0.3$, we register a loss of precision of the inference, with fluctuating mean values and wide standard deviations. Fluctuations in the estimate of the number of DOFs were also seen in the study of networks of Rossler oscillators in the work of Hahene and are likely related to nonlinear dynamics. Although preliminary, these compelling results on the effect of nonlinearities in a 1D mass-spring lattice support the model-agnostic value of the detection matrix as an inference tool. 
}

\section{Application to trusses}
\label{s:truss}

We now demonstrate the applicability of the method described thus far to  {trusses} of various degree of complexity. The simulations carried out herein rely on the same modeling considerations discussed in Section~\ref{s:sim}. The global stiffness matrix is obtained by assembling the properly-rotated stiffness matrices of each bar; the mass is considered to be distributed over the bars. Once again, we opt to only excite and measure displacements, without loss of generality.

\begin{figure*}[!htb]
\centering
\includegraphics[scale=1.1]{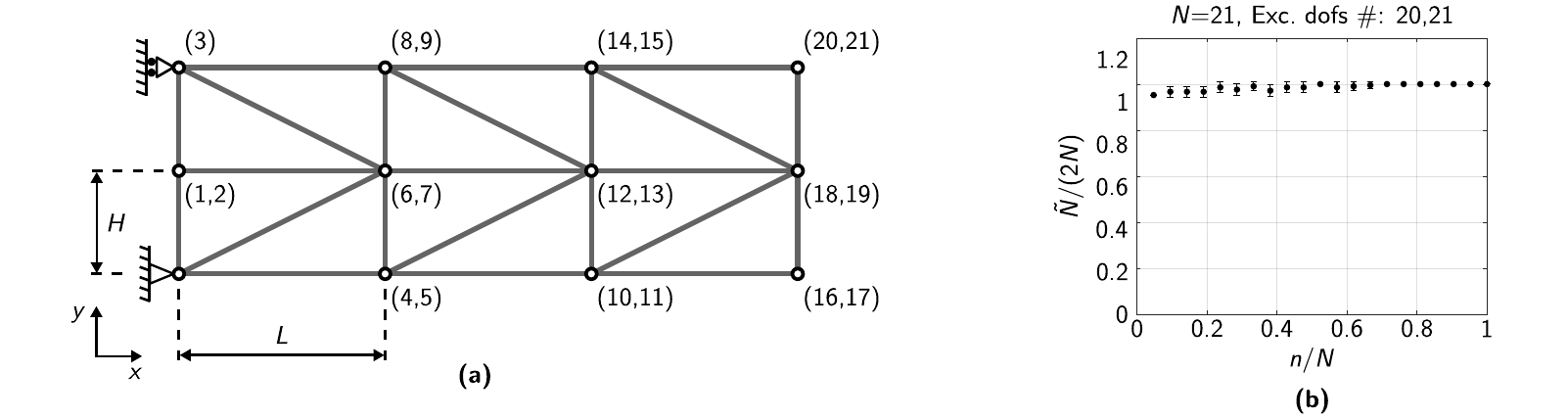}
\caption{(a) Truss from Ref.~\onlinecite{Gawronski1991}; the numbers indicate the degrees of freedom of each joint. (b) Evolution of the {$\tilde{N}/(2N)$} as a function of the number of recorded time traces, $n$, for the  {system} in (a). Each point is the average of 100 realizations and the error bars indicate the standard deviation. In each simulation, the initial conditions are only applied to degrees of freedom 20 and 21.}
\label{f:truss2}
\end{figure*}

As a first example, we choose the same truss used by Gawronski and Williams in the context of model reduction~\cite{Gawronski1991}. This is a stable, isostatic truss featuring 23 bars, 12 joints, and 21 DOFs, as illustrated in Fig.~\ref{f:truss2}a.
We choose $L=1\,\mathrm{m}$, $H=L/2$, and the cross section of each bar to be $6.25\,\mathrm{cm}^2$. The selected material is steel, with Young's modulus $E=203.4\,\mathrm{GPa}$, Poisson's ratio $\nu=0.3$, and density $\rho=8050\,\mathrm{kg\,m^{-3}}$. Note that the dimensions of the  {truss} and the selected material properties do not significantly affect our results. In this case, the generalized coordinates are the displacements of the joints along the horizontal and vertical directions.
We apply initial displacements to degrees of freedom 20 and 21, corresponding to the top-right corner of the  {truss}. We then record a varying number $n$ of time traces, from 1 to 21. Each simulation is carried out for a number of time instants corresponding to $S\Delta t/T_0=1592$, and we choose $P=72$. For each choice of $n$, we perform 100 simulations and record averages and standard deviations of {$\tilde{N}/(2N)$}, depicted in Fig.~\ref{f:truss2}b. {From our results, we determine that, while recording up to four time trace (up until $n/N=0.2$) yields a slight underestimation}, any larger value yields a very accurate estimation of the number of DOFs of the system. 

\begin{figure*}[!htb]
\centering
\includegraphics[scale=1.1]{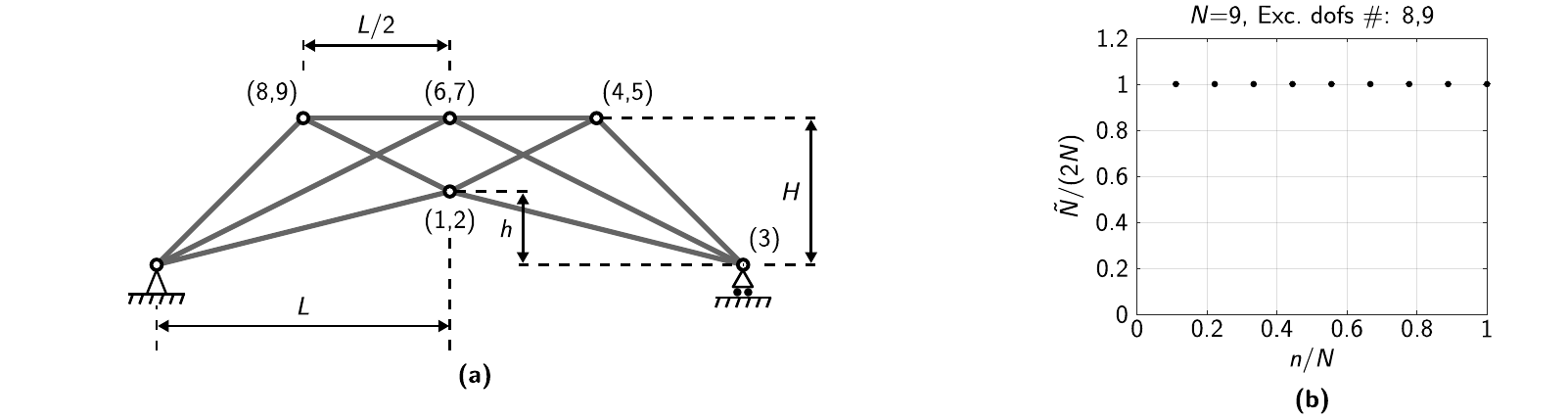}
\caption{(a) Textbook example of a complex truss with overlapping bars~\cite{Leet2008}. (b) Evolution of {$\tilde{N}/(2N)$} as a function of the number of recorded time traces, $n$, for the system in (a). Each point is the average of 100 realizations and the error bars indicate the standard deviation. In each simulation, the initial conditions are only applied to degrees of freedom 8 and 9.}
\label{f:truss1}
\end{figure*}

As a second example, we select a ``complex'' truss featuring some overlapping bars~\cite{Leet2008}. This system is shown in Fig.~\ref{f:truss1}a. In this case, we choose $L=1\,\mathrm{m}$, $H=L/2$, and $h=L/4$, while the remaining geometrical quantities, material properties and simulation parameters are the same as in the previous example.
In this case, the main challenge is posed by the topological complexity of the system, that features bars connecting non-nearest-neighbor joints. We apply initial displacements at DOFs 8 and 9, and we once again vary $n$. The topological complexity does not affect our method. In fact, we can successfully infer the correct number of DOFs in the system by measuring a single time trace.

These examples highlight that our method can be applied to complex engineering structures. One needs to keep in mind that randomizations of the system's properties will improve the performance of our method, while the presence of damping {and noise} might challenge the accuracy of the estimates -- as discussed in the context of 1D lattices. { With regards to damage detection in trusses, we acknowledge a number of challenges that remain to be addressed. In fact, in trusses, the detection matrix constructed from nodal displacements and velocities will only show signatures of damage if it causes one of the nodal locations to be isolated from the rest of the system. In this sense, it might be more convenient to develop algorithms where the detection matrix is assembled from bar strains instead of nodal quantities. We believe that this aspect deserves a separate treatment with respect to the current effort.}





\section{Conclusions}
\label{s:con}

The field of network theory could offer new pathways for solving several open questions in SHM, where there is a dire need of model-agnostic techniques to support diagnostics from time-series. Here, we make a first, necessary step in this direction by demonstrating the possibility of applying the notion of detection matrix, originally developed by Haenhe \textit{et al.}~\cite{Haehne2019}, to infer the number of DOFs of a  {mechanical system or engineering structure} from  time-series of only a small subset of its DOFs. In particular, we showed that access to time-series of as few as a single node is sufficient for the exact inference of the number of DOFs in a  {system}. 

We  {analytically} demonstrated that the approach never overestimates the number of DOFs and yields exact inferences under broad conditions on the quality and quantity of measurements  {for linear dynamics without noise}. Our results build upon the control-theoretic framing of the detection matrix that was established by Porfiri \cite{Porfiri2020}, with two {  technical differences. First, we considered the case of structural dynamics, which takes the form of a second order system of differential equations. Second, we used a single experiment to construct the detection matrix}. Finally, we offered preliminary evidence that the singular values of the detection matrix encode valuable information about potential damage in the  {system}, upon which SHM algorithms could be formulated. 

The present study is not free of limitations, which call for several lines of future inquiry. First, general criteria for the partitioning of the time-series are yet to be formulated. It is tenable that the theory of generalized Vandermonde matrices would help formulate such criteria, but this is an open area of investigation \cite{sobczyk2002generalized}. Second, general conditions for observability in the presence of non-proportional damping are elusive, thereby challenging the physical interpretation of the hypotheses regarding the applicability of the detection matrix.  Third, numerical results presented thus far only entail isostatic systems, thereby calling for further research on hyperstatic  {engineering} structures, bending-dominated frames, and random beam lattices.

When it comes to comparison with other SHM methods, a method based on the detection matrix would be quite unconventional. Unlike other parameters such as resonance frequencies and mode shapes, the rank of the detection matrix would not be affected by environmental factors, thus bypassing common issues affecting these methods. In addition, the absence of a background model to be identified constitutes a major advantage of the method, which, in principle, could be extended to include nonlinear dynamics along a similar line of work as explored by Haenhe \textit{et al.}~\cite{Haehne2019}. Beyond SHM, we envision that the detection matrix method could be used at smaller scales to evaluate the presence of damage or missing/broken connections in beam networks and cellular solids.

\section*{Acknowledgements}
PC acknowledges the support of the Research Foundation for the State University of New York. MP acknowledges the support of the National Science Foundation under grant number CMMI 1932187.

\section*{Data Availability Statement}
The data that support the findings of this study are available from the corresponding author upon reasonable request.

\appendix
\setcounter{figure}{0} 

{
\section{Equivalence between the detection matrix and a Hankel matrix}
\label{a:equivalence}

In this Appendix, we demonstrate the equivalence between the continuous-time, network-theoretic concept of detection matrix and a block Hankel matrix for discrete-time systems. To ease this comparison, we make an effort to adapt as much as possible the notation of the widespread textbook by van Overschee and de Moor~\cite{van2012subspace}, compatibly to what we have presented thus far. Therein, the authors consider a forced discrete system with null initial conditions, from which they construe a methodology to identify a system realization from knowledge of the input and the output. Borrowing the notation presented therein, we use $k$ to denote the discrete time-step, $\mathbf{x}^d_k$ the state, $\mathbf{u}_k$ the input, $\mathbf{y}_k$ the output, $\mathbf{A}$ the state matrix, $\mathbf{B}$ the input matrix, and $\mathbf{C}$ the measurement matrix (we exclude the feed-forward matrix, for simplicity). Overall, the discrete-time model is
    \begin{equation}
       \mathbf{x}^d_{k+1}=\mathbf{A}\mathbf{x}^d_{k}+\mathbf{B}\mathbf{u}_{k},
    \end{equation}
with output given by
\begin{equation}
       \mathbf{y}^{k}=\mathbf{C}\mathbf{x}^d_{k}.
    \end{equation}

In order to specialize the treatment by van Overschee and de Moor to the transient response that is encoded by a detection matrix, we assume that $\mathbf{B}=\mathbf{I}_{d\times d}$ (with $d$ being the state dimension) and that the input takes the form $\mathbf{u}_k=0$ except of $\mathbf{u}_{i-1}=\mathbf{Ax}_0$, where $i$ is a chosen time-step to partition the time axis into ``past'' and ``future,'' and $\mathbf{x}_0$ is some initial condition that we  include to recover a transient response, needed for the definition of a detection matrix.  
By utilizing the notation in Section 2.1.2 in Ref.~\onlinecite{van2012subspace}, we introduce the input and output block Hankel matrices $\mathbf{U}_p\in\mathbb{R}^{id\times j}$, $\mathbf{U}_f\in\mathbb{R}^{id\times j}$, $\mathbf{Y}_p\in\mathbb{R}^{in\times j}$, and $\mathbf{Y}_f\in\mathbb{R}^{in\times j}$, defined entry-wise as follows: 
\begin{equation}
    \mathbf{U}_p=
\left[\begin{array}{cccccc} 
    \mathbf{0}_d & \mathbf{0}_d & \mathbf{0}_d & \cdots & \mathbf{0}_d & \mathbf{0}_d \\
\vdots & \vdots & \vdots & \ddots & \vdots & \vdots \\
\mathbf{0}_d & \mathbf{0}_d & \mathbf{0}_d & \cdots & \mathbf{0}_d & \mathbf{Ax}_0
\\
\mathbf{0}_d & \mathbf{0}_d & \mathbf{0}_d & \cdots & \mathbf{Ax}_0 & \mathbf{0}_d
\\
\vdots & \vdots & \vdots & \ddots & \vdots & \vdots \\
\mathbf{Ax}_0 & \mathbf{0}_d & \mathbf{0}_d & \cdots & \mathbf{0}_d & \mathbf{0}_d
\end{array}
\right],\quad\mathbf{U}_f=\mathbf{0}_{id\times j},
\end{equation}
and
\begin{multline}
    \mathbf{Y}_p=
\left[\begin{array}{ccccc} 
    \mathbf{y}_0 & \mathbf{y}_1 & \mathbf{y}_2 & \cdots  & \mathbf{y}_{j-1} \\
    \mathbf{y}_1 & \mathbf{y}_2 & \mathbf{y}_3 & \cdots  & \mathbf{y}_{j}\\
\vdots & \vdots & \vdots & \ddots & \vdots \\
\mathbf{y}_{i-1} & \mathbf{y}_i & \mathbf{y}_{i+1} & \cdots & \mathbf{y}_{i+j-2}
\end{array}
\right],\\ 
\mathbf{Y}_f=
\left[\begin{array}{ccccc} 
    \mathbf{y}_i & \mathbf{y}_{i+1} & \mathbf{y}_{i+2} & \cdots  & \mathbf{y}_{i+j-1} \\
    \mathbf{y}_{i+1} & \mathbf{y}_{i+2} & \mathbf{y}_{i+3} & \cdots  & \mathbf{y}_{i+j}\\
\vdots & \vdots & \vdots & \ddots & \vdots \\
\mathbf{y}_{2i-1} & \mathbf{y}_{2i} & \mathbf{y}_{2i+1} & \cdots & \mathbf{y}_{2i+j-2}
\end{array}
\right],
\end{multline}
where subscripts $p$ and $f$ indicate past and future, $2i+j-1$ is the total number of available measurements from an initial time $t=0$ (similar to $S$ in our continuous-time formulation), and the case chosen for illustration purposes is $j<i$. 

The key matrix that is used to determine the size of the system $d$ from Theorem 2 in Section 2.2.2 of Ref.~\onlinecite{van2012subspace} is 
\begin{equation}
    \mathbf{\mathcal{O}}_i=\mathbf{Y}_f/_{\mathbf{U}_f}\mathbf{W}_p
\end{equation}
where $\mathbf{W}_p=[\mathbf{U}_p^{\mathrm{T}},\mathbf{Y}_p^{\mathrm{T}}]^{\mathrm{T}}$ and the symbol ``/'' identifies the so-called oblique projection as defined in Section 1.4.2 of Ref.~\onlinecite{van2012subspace}. Under quite general conditions on all the involved quantities, it can be shown that the rank of this matrix is equal to $d$\footnote{Note that the conditions of Theorem 2 require persistence of excitation that is obviously not satisfied with the chosen input that leads to a rank-deficient Hankel matrix; however, such a condition is not needed for all the claims of Theorem 2, in particular for the one we are interested in on the determination of $d$.}. For the chosen input sequence, we  {show} that such a matrix corresponds to a detection matrix like the one we have employed in this study. Specifically, we show that 
    \begin{equation}
    \mathbf{\mathcal{O}}_i=\mathbf{Y}_f,
    \end{equation}
which posits that the main matrix used in Theorem 2 can be obtained by aggregating the time-series of the transient response of the system, measured from its output. 

To  {derive} such a claim, we begin applying Corollary 1 in Section 1.4.2 of Ref.~\onlinecite{van2012subspace}, by which the oblique projection can be simplified into an orthogonal projection due to matrix $\mathbf{U}_f$ being zero. As such, we can write 
\begin{equation}\label{eq:orthp}
    \mathbf{\mathcal{O}}_i=\mathbf{Y}_f/\mathbf{W}_p.
\end{equation}
From the steps in the proof of Theorem 2 (combination of equations (2.15) and (2.16) in Ref.~\onlinecite{van2012subspace}) and the fact that $\mathbf{U}_f$ is zero for the case at hand, we establish
    \begin{equation}\label{eq:YfWp}
        \mathbf{Y}_f=\mathbf{Q}\mathbf{W}_p,
    \end{equation}
where matrix $\mathbf{Q}$ is related to the observability matrix and powers of $\mathbf{A}$. Based on \eqref{eq:YfWp}, the rows of matrix $\mathbf{Y}_f$ are in the row space of $\mathbf{W}_p$, so that the orthogonal projection in \eqref{eq:orthp} returns the same matrix. Differences between the ordering and the sampling of the elements of $\mathbf{Y}_f$ with respect to the detection matrix introduced for the study of second-order systems is a minor difference, which bears no implication with respect to the rank computation. Of interest, as mentioned in our Section~\ref{s:equivalence}, is instead the fact that our original theorem applies to any class of continuous-time systems, without the need of introducing a discretization of the model.
}

%

\end{document}